\documentclass[runningheads]{llncs}
\usepackage{booktabs} 
\usepackage{wrapfig}
\usepackage{pgfplots}
\usepackage{float}
\usepackage{graphicx}
\usepackage{textcomp}
\usepackage{multirow}
\usepackage{multicol}
\usepackage{caption}
\usepackage{wrapfig}
\usepackage{tikz}
\usepackage{amsmath}

\begin{document}
\title{Hardware realization of residue number system algorithms by Boolean functions minimization}
\titlerunning{Hardware realization of arithmetic operations}
%
\author{Danila Gorodecky\inst{1} \and Tiziano Villa\inst{2}}
\authorrunning{D. Gorodecky and T.Villa}
%
\institute{National Academy of Science of Belarus, Minsk, Belarus \\ \email{danila.gorodecky@gmail.com}\\ \and
University of Verona, Verona, Italy \\
\email{tiziano.villa@univr.it}}
\maketitle              
\begin{abstract}
Residue number systems (RNS) represent numbers by their remainders modulo a set of relatively prime numbers. This paper proposes an efficient hardware implementation of modular multiplication and of the modulo function ($X(mod\ P)$), based on Boolean minimization. 
We report experiments showing a performance advantage up to 30 times for our approach vs. the results obtained by state-of-art industrial tools.

\end{abstract}
\section{Introduction}

The idea of the Residue Number System (RNS) goes back to an ancient Chinese source showing how to convert residues into numbers, and was later formalized by C.F. Gauss in the 19th century. Since the advent of digital computers, there have been many papers proposing algorithms to implement efficiently RNS on computers.

The main advantage of RNS is the speed and reliability of arithmetic computations \cite{cher,so_ch,zimm}. The first application of RNS was in the search of prime numbers. Nowadays implementations of RNS can be found in anti-aircraft systems \cite{malash}, neural computations \cite{cher}, real-time signal processing (pattern recognition) \cite{fl_he_fl_pi}, cryptography \cite{oz_su_sa}. Modular arithmetic (MA) is effective for processing large data flows (with several hundreds or thousands bits) \cite{mont}. 

RNS is a form of parallel data processing, where computer arithmetic is performed using the residues of the division by a pre-selected base of co-primes moduli $\{p_1,p_2,...,p_m\}$. The residues have a lower number of digits than the original numbers and arithmetic operations over the residues can be performed separately for each modulo of the base, resulting in faster processing (e.g., faster addition and multiplication), compared to other forms of parallel data processing.

\begin{figure}
\includegraphics[height=6cm, width=1\linewidth]{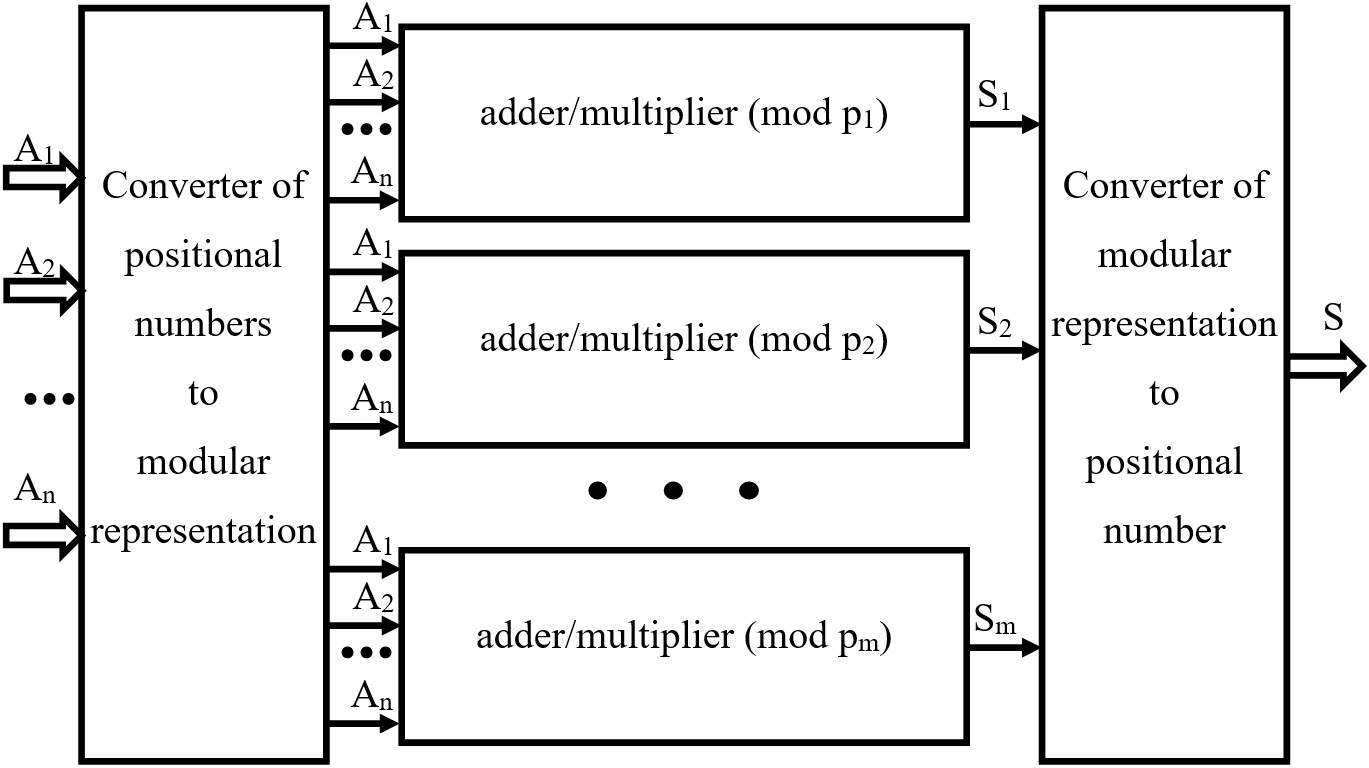}
\caption{Common structure of RNS.}
\label{fig:RNS}
\end{figure}

Data processing in modular arithmetic includes the following steps. Firstly, input operands $A_1, A_2, \dots, A_n$ are converted from positional to modular representations computing the remainders (or residues) with respect to the moduli $\{p_1, p_2, \dots, p_m\}$ (see left block in Fig.~\ref{fig:RNS}); then arithmetic operations over the residues of the operands for each modulo $\{p_i\}$, where $i =1, \dots, n$, are computed (middle block in Fig.~\ref{fig:RNS}); finally, the results $S_1,S_2,...,S_m$ for each modulo are converted back from modular to positional representations $S$ (see right block in Fig.~\ref{fig:RNS}).

Conversion into modular representation (direct conversion) is realized by the modulo $X (mod\ P)$ function, whose result is fed into the second step of operations.
The second step of the RNS computation requires performing modular summation, multiplication, and other arithmetic functions such $A \cdot B + C$.
The third step in RNS computes the polynomial form $S_1 \cdot C_1 + S_2 \cdot C_2 + ... + S_m \cdot C_m - P \cdot r$, where $S_1, S_2, ...$ are outputs of the previous step, $C_1, C_2,...$ are pre-calculated constants, $r$ is a constant which is obtained during the computation of the polynomial, and $P=p_1\cdot p_2\cdot ... \cdot p_m$. In other words  the third step in RNS computes $(S_1 \cdot C_1 + S_2 \cdot C_2 + ... + S_m \cdot C_m) (mod ~P)$. Therefore, the main arithmetic operations needed for RNS computations are the modulo function $X (mod\ P)$, modular summation, and modular multiplication.

A major limitation when processing large numbers in RNS is the complexity of hardware realization of converters (left and right blocks in Fig.~\ref{fig:RNS}). This is due to the fact that to compute the modulo function and to recover the positional representation one should perform division, modular multiplication, and comparison. There are different approaches to solve this problem (e.g. \cite{cher,zimm,comp}), but, mostly, they are restricted with respect to the modular values (e.g., $mod\ 2^k-1, mod\ 2^k, mod\ 2^k+1$) and to the number of operands.

In this paper we describe algorithms for the modulo function ($X (mod\ P)$) and for modular multiplication. We report experimental results comparing with industrial tools (Synopsys and Mentor Graphics).

\section{Hardware design of arithmetic units}

The approach that we propose is characterized as follows:
\begin{enumerate} 
\item It is valid for an arbitrary modulo and bit range of the inputs. 
\item It can be applied to modular multiplication and to the modulo function.
\item It is based on combinational logic.
\end{enumerate}
In the literature we can find techniques to compute the modular multiplication \cite{om_pr} and the modulo function \cite{comp,om_pr}, but they are based on memory usage and require a big area with high latency.
In the proposed procedures, there are some common tasks:
\begin{enumerate}
\item Inputs (input factors $A \cdot B$ in multiplication or input $X$ in $X (mod\ P)$) are split into subvectors.
\item All subvectors are combined to define a polynomial.
\item This procedure is iterated as long as the result $> 2 \cdot P$.
\end{enumerate}

\subsection{Computation of the modulo function}
Modulus function $X (mod\ P)$ can be computed by means of combinational or sequential circuits.

Some sequential realizations store pre-calculated values of modulus function by \cite{moh,irh}, computed by using an automaton model \cite{will_ryan}, or resort to pipelining using a chain of homogeneous arithmetic blocks \cite{bu_sa}, where every term corresponds to an arithmetic block in hardware:
\begin{equation*}\label{XmodP}
\begin{split}
X & = P\cdot Q + R \\
 & = P\cdot 2^{\delta} \cdot q_{\delta} + P\cdot 2^{\delta -1} \cdot q_{\delta -1} + \ldots + P\cdot 2^0 \cdot q_0 + R
\end{split}
\end{equation*}
and $X (mod\ P) = R$, where $X=(x_{\psi},x_{\psi -1},\ldots,x_1)$ and $\delta$ is defined by the inequality $P\cdot 2^{\delta +1} < 2^{\psi}-1 \leq P\cdot 2^{\delta}$. Notice that $P$ can be an arbitrary number.

Approaches with no memory that are efficient with respect to performance and area require special moduli sets \cite{om_pr}, which consist of variations of $2^s \pm v$, where $v=1,3,5$: $\{2^s-1,2^s,2^s+1\}$, $\{2^{2\cdot s}-1,2^s,2^{2\cdot s}+1\}$, $\{2^{s}-1,2^{2\cdot s},2^{s}+1,2^{s-1}-1,2^{s+1}-1\}$, etc.

Given that in the RNS representation the moduli must be co-prime numbers, multiplication of two 1000-bit numbers using the moduli $\{2^s-1,2^{2\cdot s},2^s+1,2^{s-1}-1, 2^{s+1}-1\}$ requires $s \approx 400$ bits, which impairs the computational efficiency of the transformation. The same multiplication can be realized using a set of smaller moduli, since there are more than 400 up to 12-bit numbers that are co-prime. Note that in order to represent numbers in RNS uniquely the result of the calculation must not exceed $P=p_1\cdot p_2\cdot ... \cdot p_m$. If $P=(2^s-1)\cdot (2^{2\cdot s})\cdot (2^s+1) \cdot (2^{s-1}-1)\cdot (2^{s+1}-1)$, then $s$ takes approximately 400-bit number.

We propose the following two-step procedure to compute $X (mod\ P)$:
\begin{enumerate}
\item
$X$ is split into $k$ subvectors with $\leq \delta$ bits in every subvector, where $\delta = [log_2{P-1}]$.
\item
The resulting subvectors are combined according to Eq~\ref{eq1}:
\begin{equation}
\label{eq1}
X (mod\ P)=\sum\limits_{i=1}^k {X_i \cdot \big(2^{\delta \cdot (i-1)} (mod\ P) \big)}.
\end{equation}
\end{enumerate}
This formula can be applied recursively producing reduced intermediate results at every step. The coefficient $2^{\delta \cdot (i-1)}(mod\ P)$ is a constant and it does not exceed $P-1$. At the first step, it holds that $X_i = 2^\delta-1$, since Eq.~\ref{eq1} achieves the maximum value. Then Eq.~\ref{eq1} is called recursively until the result is $\leq 2 \cdot P$. At the end, the result is compared with $P$ and, if needed, $P$ is subtracted from the result of the last step. 

For illustration, consider the following example. Suppose that $X$ is an 18-bit input and $P = 47$. Then modulo $P$ is a 6-bit number, and the input $X$ is split into three 6-bit tuples $X = (X_3, X_2, X_1)$, where $X_1 = (x_6,x_5, \dots, x_1)$, $X_2 = (x_{12},x_{11}, \dots, x_7)$, and $X_3 = (x_{18},x_{17}, \dots, x_{13})$. Then $2^6(mod\ 47) = 17 (mod\ 47)$ and $2^{12}(mod\ 47) = 7 (mod\ 47)$. Hence, in the first iteration Eq.~\ref{eq1} takes the following form:
\begin{equation*}
\begin{split}
X (mod\ 47) & = X_1 + X_2 \cdot 2^6(mod\ 47) + X_3 \cdot 2^{12}(mod\ 47)=  \\
& = X_1 + X_2 \cdot 17 (mod\ 47) + X_3 \cdot 7 (mod\ 47) = S_1
\end{split}
\end{equation*}
If input $X = 2^{18}-1$, then its binary representation requires 18 bits, i.e., $X_1 = X_2 = X_3 = 63_{10} = 111111_{2}$. Then $S_1 \leq 63 + 63 \cdot 17 + 63 \cdot 7 = 1575_{10} = 11000100111_{2}$. In this case Eq.~\ref{eq1} takes the following form:
\begin{equation*}
\begin{split}
S_1 (mod\ 47) & = S_1^1 + S_2^1 \cdot 2^6(mod\ 47) =  \\
& = S_1^1 + S_2^1 \cdot 17(mod\ 47) = S_2 \leq 447.
\end{split}
\end{equation*}
If $S_1^1=1001110_{2}$ and $S_2^1=11000_{2}$, it follows $S_2 = 447$. The second iteration splits the 9-bit $S_2$ number into two 6-bit and 3-bit tuples: $S_2=\big( S_2^2,S_1^2\big)$, where $S_2^2 = \big(s_9^2,s_8^2,s_7^2\big)$ and $S_1^2 = \big(s_6^2,s_5^2,...,s_1^2\big)$. In this case Eq.~\ref{eq1} takes the following form:
\begin{equation*}
\begin{split}
S_2 (mod\ 47) & = S_1^2 + S_2^2 \cdot 17 (mod\ 47) = S_3 \leq 148.
\end{split}
\end{equation*}
If $S_1^2=111111_{2}$ and $S_2^2=101_{2}$, it follows $S_3=148$. The third iteration splits the 8-bit number $S_3$ into two 6-bit and 2-bit tuples: $S_3=\big( S_2^3,S_1^3\big)$, where $S_2^3 = \big(s_8^3,s_7^3\big)$ and $S_1^3 = \big(s_6^3,s_5^3,...,s_1^3\big)$. In this case Eq.~\ref{eq1} takes the following form:
\begin{equation*}
\begin{split}
S_3 (mod\ 47) & = S_1^3 + S_2^3 \cdot 17 (mod\ 47) = S_4 \leq 54.
\end{split}
\end{equation*}
If $S_1^3=010100_{2}$ and $S_2^3=10_{2}$, it follows $S_4=54$.
Since $S_4<2 \cdot P = 94$, $S_4$ is compared with $P = 47$: if $S_4>47$, then $X (mod\ 47) = S_4-47$, else $X (mod\ 47) = S_4$.

\subsection{Computation of the modular product}

We propose the following two-step procedure to compute the product
$A \cdot B = R (mod\ P)$, where $A=(A_{\delta},A_{\delta-1},...,A_1)$, 
$B=(B_{\delta},B_{\delta-1},...,B_1)$, and the $\delta$-subvectors $A_{\delta}$ and $B_{\delta}$ consist of the most significant bits. For example, if $A$ and $B$ are 12-bit numbers and $\delta=4$ , then $A_4=(a_{12},a_{11},a_{10})$ and $B_4=(b_{12},b_{11},b_{10})$, where $a_{12}$ and $b_{12}$ are the most significant bits.

This contribution proposes a modulus function computation for an arbitrary modulo without limitation on the value of $P$. The idea of the approach is to use a large set of small moduli vs. a small set of large moduli, as it is used traditionally. Hence we consider that $A,B$ and $P$ vary from 6 to 12 bits. 
\begin{enumerate}
\item
The inputs are split into 2-, 3- and 4-bit subvectors.
\item
The corresponding pairs of subvectors are multiplied applying the following recursive formula:
\begin{equation}
\label{eq2}
R=\sum\limits_{i=1}^{\delta}  \sum\limits_{j=1}^{\delta} {A_i \cdot B_j \cdot \big( 2^{m \cdot (i+j-2) \cdot 3}} (mod\ P) \big)=S\_temp,
\end{equation}
\end{enumerate}
The maximum value of $S\_temp$ does not exceed $2^{3 \cdot \delta +2}$, $2^{3 \cdot \delta +3}$ or $2^{3 \cdot \delta +4}$ depending on value of modulo $P$. 

As an illustration, consider three common cases:
\begin{enumerate}
\item $\delta = 2$, then $S\_temp \leq 2^8$ and $S\_temp_2= S \_temp[3:1] + S \_temp[6:4] \cdot 2^3(mod\ P) + S \_temp[8:7] \cdot 2^6 (mod\ P)$;
\item $\delta = 3$, then $S\_temp \leq 2^{12}$ and $S\_temp_2= S \_temp[3:1] + S\_temp[6:4] \cdot 2^3(mod\ P) + S \_temp[9:7] \cdot 2^6 (mod\ P) + S\_temp[12:10] \cdot 2^9 (mod\ P)$;
\item $\delta = 4$, then $S\_temp \leq 2^{12}$ and $S\_temp_2 = S\_temp[3:1] + S\_temp[6:4] \cdot 2^3 (mod\ P) + S \_temp[9:7] \cdot 2^6 (mod\ P) + S\_temp[12:10] \cdot 2^9 (mod\ P) + S \_temp[15:13] \cdot 2^{12} (mod\ P)$.
\end{enumerate}
Finally, if $S\_temp_2 > P$, then $S=S\_temp_2 - P$, otherwise $S=S\_temp_2$.

Let us multiply the two 6-bits numbers $A \cdot B = S (mod\ 47)$. Splitting operands into two, i.e., $\delta = 2$, 3-bits subvectors, Eq.~\ref{eq2} is transformed in the following form:\\
$A \cdot B = S (mod\ 47) = A_1 \cdot B_1 (mod\ 47) + A_1 \cdot B_2 \cdot 2^3 (mod\ 47) + A_2 \cdot B_1 \cdot 2^3 (mod\ 47) + A_2 \cdot B_2 \cdot 2^6 (mod\ 47) = S\_temp$.

When $A = 45$ and $B = 15$, $S\_temp$ achieves the maximum value, which is $158_{10} = 10011110_{2}$: $A_1=101_2$, $A_2=101_2$, $B_1=111_2$, $B_2=1_2$, hence $A \cdot B = 5\cdot 7 (mod\ 47) + 5 \cdot 1 \cdot 2^3 (mod\ 47) + 5 \cdot 7 \cdot 2^3 (mod\ 47) + 5 \cdot 1 \cdot 2^6 (mod\ 47) = 35 (mod\ 47) + 40 (mod\ 47) + 45 (mod\ 47) + 38 (mod\ 47) = 158$. Trying another value for $A$ and $B$, it is $S\_temp < 158$.

The second iteration reduces $S\_temp$ to a value $<47$. Assume that $S\_temp=158$, then
$S\_temp_2 = 6 + 3 \cdot 2^3 (mod\ 47) + 2 \cdot 2^6 (mod\ 47) = 6 + 24 + 34 = 64$.

Finally, taking into account that $64>47$, the result is $S=64-47=17$.
Note that the bit range of $S\_temp$ is preselected.

\section{Boolean minimization in modular operations}
The result of any arithmetic computation can be represented as sum-of-products (SOPs). However the original representation given by truth tables may be unmanageable  by synthesis tools, e.g., the truth table of the product of two 16-bit input operands requires 64 columns (16 columns for each operand and 32 columns for the result) and more than four billions rows.

For a pair of $\delta$-bit tuples, consider $2^i(mod\ P)$ $X_i\cdot\big( 2^{\delta\cdot(i-1)}(mod\ P)$, where $i=1,2, \dots, k$, are the corresponding factors of the multiplication. Then $2^i(mod\ P)$ is a constant whose bits are redundant in the minimization, because all rows in the truth table corresponding to this constant have the same value $2^i(mod\ P)$.

The initial truth table for $X (mod\ P)$ consists of $P$ rows and $2 \cdot \delta$ columns, where the left $\delta$ columns correspond to all integers from $0$ up to $P-1$, and the right columns correspond to $X\cdot 2^i(mod\ P)$. 

{\bf Example} Consider $2^8(mod\ 13)=9(mod\ 13) = 1001_{2}$. In this case, subtable 1 represents the truth table for $X\cdot 9(mod\ 13)$ before minimization and subtable 2 represents the SOP after minimization (it can obtained by tools like~\cite{espresso} or ELS~\cite{b_c_k_k_r_c}).
So the first four bits in the last row of the truth table in subtable 1 represent $12_{10}$, and the right four bits represent $12\cdot 9(mod\ 13)=4_{10}$.
For the 18-bit input $X$ and $P = 47$, all pairs of corresponding factors are represented as a SOP: with 12 columns (6 inputs and 6 outputs) $X_2\cdot 17(mod\ 47)$ and $X_3\cdot 7(mod\ 47)$; with 11 columns (5 inputs and 6 outputs) $X_2^1\cdot 17(mod\ 47)$; with 9 columns (3 inputs and 6 outputs) $X_2^2\cdot 17(mod\ 47)$; with 8 columns (2 inputs and 6 outputs) $X_2^3\cdot 17(mod\ 47)$.

\begin{center}
\begin{table}\caption{Representation of $X (mod\ 13)$ with SOPs}
\begin{center}
\begin{tabular}{ |c c|c c| } 
$a_2 a_1 b_2 b_1$ & $r_4 r_3 r_2 r_1$ & $a_2 a_1 b_2 b_1$ & $r_4 r_3 r_2 r_1$ \\
\hline
0 0 0 1 & 0 0 0 0 & 0 1 0 0 & 1 0 0 0 \\
0 0 0 0 & 1 0 0 1 & 1 0 1 - & 1 0 0 0 \\
0 0 1 0 & 0 1 0 1 & 0 0 0 1 & 1 0 0 0 \\
0 0 1 1 & 0 0 0 1 & 0 1 1 1 & 1 0 0 0 \\
0 1 0 0 & 1 0 1 0 & 0 1 0 1 & 0 1 0 0 \\
0 1 0 1 & 0 1 1 0 & - 0 1 0 & 0 1 0 0 \\
0 1 1 0 & 0 0 1 0 & 1 - 0 0 & 0 1 0 0 \\
0 1 1 1 & 1 0 1 1 & 1 0 0 - & 0 0 1 0 \\
1 0 0 0 & 0 1 1 1 & 0 1 - - & 0 0 1 0 \\  
1 0 0 1 & 0 0 1 1 & - 0 0 1 & 0 0 0 1 \\
1 0 1 0 & 1 1 0 0 & 1 0 0 - & 0 0 0 1 \\
1 0 1 1 & 1 0 0 0 & 0 - 1 1 & 0 0 0 1 \\
1 1 0 0 & 0 1 0 0 & 0 0 1 - & 0 0 0 1 \\
\cline{1-4} 
\multicolumn{2}{c}{Subtable 1} & \multicolumn{2}{c}{Subtable 2}
\end{tabular}
\end{center}
\end{table}
\end{center}

\section{Experimental results}
We compared our procedure with respect to three electronic design automation (EDA) tools: Synopsys, Mentor Graphics (for standard cells), and Xilinx (for FPGAs). Since Mentor Graphics and Xilinx do not synthesize modular operations, we compared with special moduli, such $2^{s}-1, 2^{s}+1$. Our approach shows minor gains within 10\%. 

Synopsys is the only EDA tool which generates $X (mod\ P)$ circuits. We report results of synthesis using Synopsys 2014 on 28 nm Standard Cell ASIC technology from United Microelectronics Corporation. Plots \ref{gr_1} a) and \ref{gr_1} b) compare the latency of circuits of $X (mod\ P)$ (in MHz) for inputs $X$ of 400 and 500 bits, and for moduli $P$ of 10, 11, and 12 bits, respectively. Plots \ref{gr_2} a) and \ref{gr_2} b) compare the area of circuits of $X (mod\ P)$ (number of cells from the library \textit{cells}) for inputs $X$ of 400 and 500 bits, and for moduli $P$ of 10, 11, and 12 bits, respectively.

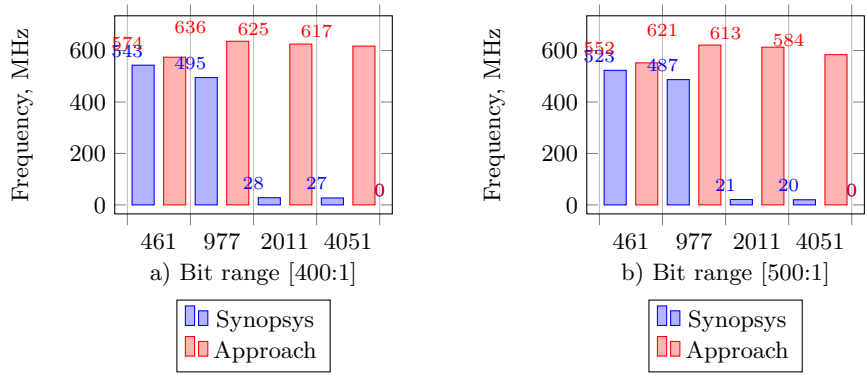
\begin{figure}[h]
\caption{\label{gr_1}Modulo function: performance comparison of our approach vs. Synopsys}
\hspace{1ex}
\begin{multicols}{2}
\begin{tikzpicture}
\begin{axis}[
    width=150pt,
    height=120pt,
	x tick label style={/pgf/number format/1000 sep=},
    symbolic x coords={461, 977, 2011, 4051, 5000},
	enlargelimits=0.05,
	legend style={ at={(0.5,-0.8)}, anchor=south},
    ymax=700,
	ybar interval=0.7,
    ylabel={Frequency, MHz},
    xlabel={a) Bit range [400:1]},
    nodes near coords,
    nodes near coords align={vertical},
    every node near coord/.append style={font=\scriptsize},
]
\addplot[bar width=5pt, blue, fill=blue!30!white] coordinates {(461,543) (977,495) (2011,28) (4051,27) (5000,0)};
\addplot[bar width=5pt, red, fill=red!30!white] coordinates {(461,574) (977,636) (2011,625) (4051,617) (5000,0)};
\legend{Synopsys, Approach}
\end{axis}
\end{tikzpicture}

\begin{tikzpicture}
\begin{axis}[
    width=150pt,
    height=120pt,
	x tick label style={/pgf/number format/1000 sep=},
    symbolic x coords={461, 977, 2011, 4051, 5000},
	enlargelimits=0.05,
	legend style={ at={(0.5,-0.8)}, anchor=south},
    ymax=700,
	ybar interval=0.7,
    ylabel={Frequency, MHz},
    xlabel={b) Bit range [500:1]},
    nodes near coords,
    nodes near coords align={vertical},
    every node near coord/.append style={font=\scriptsize},
]
\addplot[bar width=5pt, blue, fill=blue!30!white] coordinates {(461,523) (977,487) (2011,21) (4051,20) (5000,0)};
\addplot[bar width=5pt, red, fill=red!30!white] coordinates {(461,552) (977,621) (2011,613) (4051,584) (5000,0)};
\legend{Synopsys, Approach}
\end{axis}
\end{tikzpicture}
\end{multicols}
\end{figure}
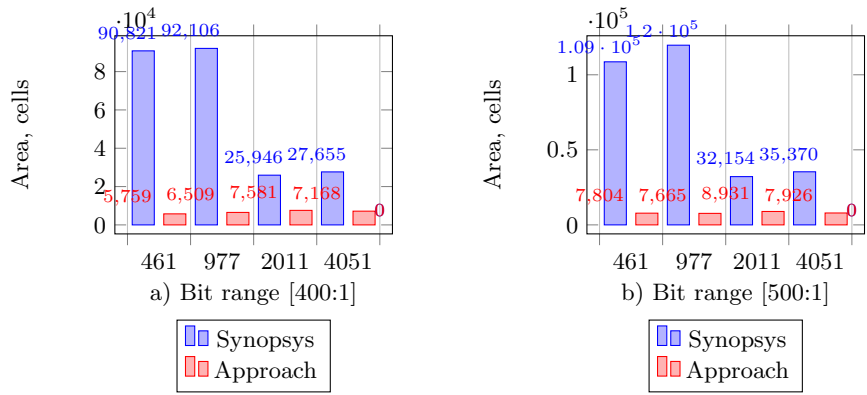
\begin{figure}[h]
\caption{\label{gr_2}Modulo function: area comparison (cells) of our approach vs. Synopsys}
\hspace{1ex}
\begin{multicols}{2}
\begin{tikzpicture}
\begin{axis}[
    width=150pt,
    height=120pt,
	x tick label style={/pgf/number format/1000 sep=},
    symbolic x coords={461, 977, 2011, 4051, 5000},
	enlargelimits=0.05,
	legend style={ at={(0.5,-0.8)}, anchor=south},
    ymax=94000,
	ybar interval=0.7,
    ylabel={Area, cells},
    xlabel={a) Bit range [400:1]},
    nodes near coords,
    nodes near coords align={vertical},
    every node near coord/.append style={font=\scriptsize},
]
\addplot[bar width=5pt, blue, fill=blue!30!white] coordinates {(461,90821) (977,92106) (2011,25946) (4051,27655) (5000,0)};
\addplot[bar width=5pt, red, fill=red!30!white] coordinates {(461,5759) (977,6509) (2011,7581) (4051,7168) (5000,0)};
\legend{Synopsys, Approach}
\end{axis}
\end{tikzpicture}

\begin{tikzpicture}
\begin{axis}[
    width=150pt,
    height=120pt,
	x tick label style={/pgf/number format/1000 sep=},
    symbolic x coords={461, 977, 2011, 4051, 5000},
	enlargelimits=0.05,
	legend style={ at={(0.5,-0.8)}, anchor=south},
    ymax=120000,
	ybar interval=0.7,
    ylabel={Area, cells},
    xlabel={b) Bit range [500:1]},
    nodes near coords,
    nodes near coords align={vertical},
    every node near coord/.append style={font=\scriptsize},
]
\addplot[bar width=5pt, blue, fill=blue!30!white] coordinates {(461,108640) (977,119715) (2011,32154) (4051,35370) (5000,0)};
\addplot[bar width=5pt, red, fill=red!30!white] coordinates {(461,7804) (977,7665) (2011,8931) (4051,7926) (5000,0)};
\legend{Synopsys, Approach}
\end{axis}
\end{tikzpicture}
\end{multicols}
\end{figure}

The experiments show significant gains by our approach compared with Synopsys. The gain in performance is up to 30 times and in area is up to 15 times.
Moreover, Synopsys could not synthesize circuits for inputs $X$ larger than 500 bits: the synthesis by Synopsys of the modulo function for a 600-bit input $X$ failed after nine days, whereas it takes only 20 minutes with our approach.

\section{Conclusions and further research}

Performance of computer arithmetic is one of the main advantages of RNS vs. traditional approaches. We proposed a technique that improves significantly area and performance of RNS with respect to synthesis using standard EDA tools.

Our approach is not limited to modular multiplication and to the modulo function, but it can be extended to any arithmetic operation. Dozens of circuits were designed with the technique presented here and then embedded in arithmetic units by the hi-tech factory Integral (Minsk, Belarus).

Topics of further research include:
\begin{enumerate} 
\item Comparing different forms of representations (SOPs, Reed-Muller expansions, binary decision diagrams) in the realization of partial products;
\item Designing FPGAs using Xilinx and Altera architectures. 

\end{enumerate} 

%
%
%
%

\end{document}